\documentclass[aps,prb,twocolumn,groupedaddress, showpacs, showkeys,floatfix]{revtex4}

\usepackage{amsmath}
\usepackage{amsfonts}
\usepackage{amssymb}
\usepackage{epsfig}

\begin{document}
\title{Entropic Stabilization and Retrograde Solubility in Zn$_4$Sb$_3$}
\author{Gregory S. Pomrehn$^1$}
\author{Eric S. Toberer$^1$}
\author{G. Jeffrey Snyder$^1$}\email[]{jsnyder@caltech.edu}
\author{Axel van de Walle$^1$}\email[]{avdw@caltech.edu}

\affiliation{1 -  Materials Science, California Institute of Technology, 1200 E. California Blvd., Pasadena, California 91125, USA}

\date{\today}
\begin{abstract}
Zn$_4$Sb$_3$ is shown to be entropically stabilized versus decomposition to Zn and ZnSb though the effects of configurational disorder and phonon free energy.   Single phase stability is predicted for a range of compositions and temperatures.  Retrograde solubility of Zn is predicted on the two-phase boundary region between Zn$_4$Sb$_3$ and Zn.  The complex temperature dependent solubility can be used to explain the variety of nanoparticle formation observed in the system: formation of ZnSb on the Sb rich side, Zn on the far Zn rich side and nano-void formation due to Zn precipitates being reabsorbed at lower temperatures.
\end{abstract}



\maketitle

\subsection{Introduction}  
The Zn-Sb binary phase system has been of interest for many years in the search for efficient and low-cost thermoelectric materials.  Of primary interest has been the Zn$_4$Sb$_3$ phase which exhibits a thermoelectric figure of merit, $zT$, in excess of 1 in intermediate temperature ranges\cite{Caillat:HighZT}.  This phase, being composed of environmentally benign and relatively earth abundant elements, continues to draw active research. 

\begin{figure}
\epsfig{file=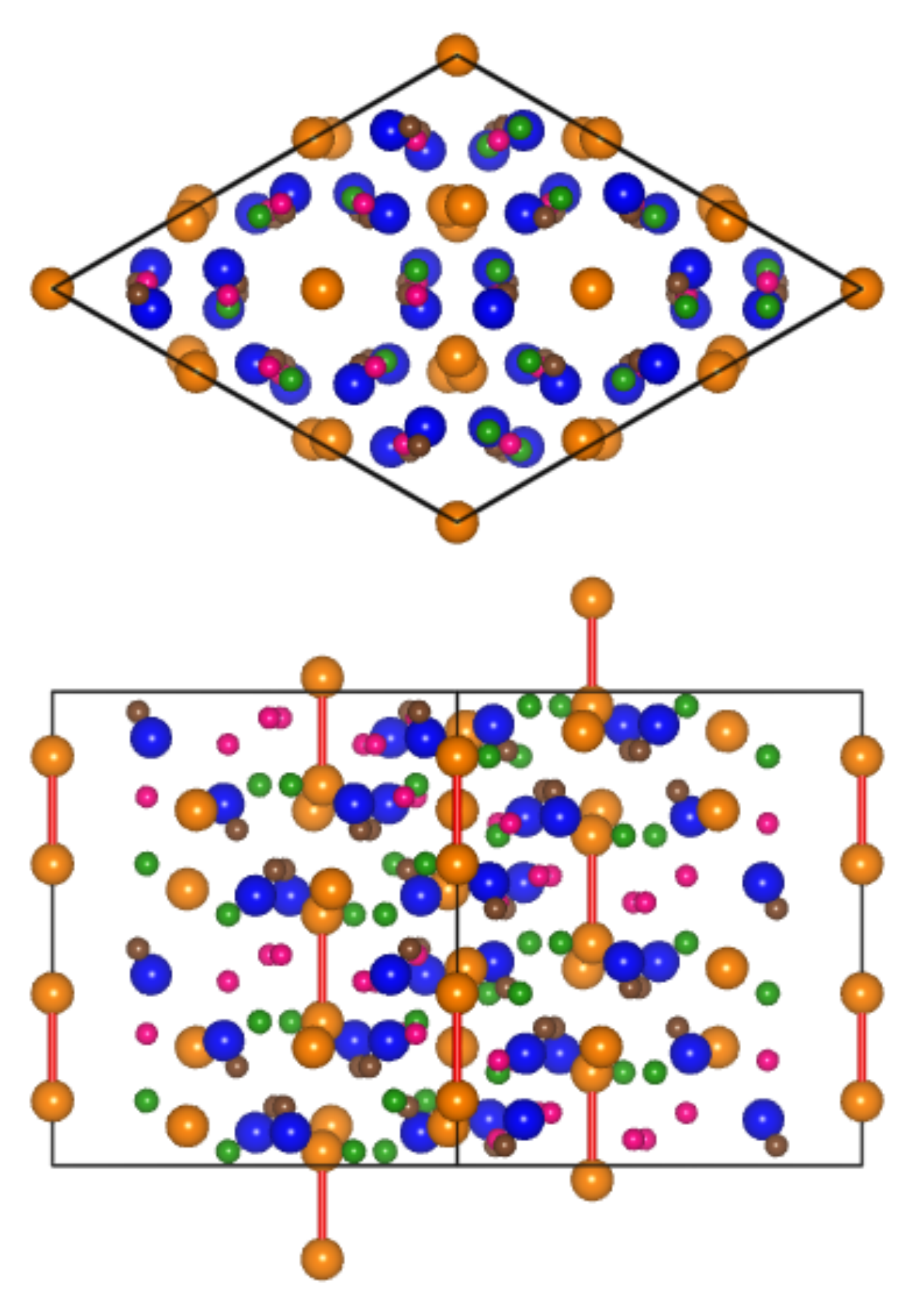, width=6.5cm}\\
\caption{(Color online) Conventional hexagonal unit cell of Zn$_4$Sb$_3$.  The Sb sub-lattice (orange) is composed of isolated Sb and Sb dimers.  The Zn `A' sub-lattice (blue) is ~90\% occupied.  The `B' (green), `C' (brown), and `D' (pink) Zn sub-lattices are ~5\% occupied.   The primitive rhombohedral cell is 1/3 the volume of the conventional cell.}
\label{structure}
\end{figure}

Zn$_4$Sb$_3$ exhibits exceptionally low lattice thermal conductivity\cite{Caillat:HighZT}, due in part to its high configurational disorder.  The room temperature structure of Zn$_4$Sb$_3$ ($R\bar{3}c$, Figure\,\ref{structure}), denoted `$\beta$-Zn$_4$Sb$_3$' here \cite{Caillat:HighZT} (or sometimes `$\epsilon$-Zn$_4$Sb$_3$'\cite{Nakamoto:Homogeneity}), contains an anionic framework composed of 30 Sb, divided between 6 Sb$_2^{4-}$ dimers and 18 isolated Sb$^{3-}$.  With 39 Zn$^{2+}$ a charge balanced composition is obtained at Zn$_{13}$Sb$_{10}$.  Of these Zn cations, at most, 36 can fit on Zn framework sites and 3 must be distributed among 3 crystallographically distinct interstitial sites\cite{SnyderNM:Zn4Sb3}.   For clarity, we will always refer to the entire phase as `Zn$_4$Sb$_3$' and write `Zn$_{X}$Sb$_{10}$' only when referring to a specific configurational composition within `Zn$_4$Sb$_3$'.

Previous ab initio studies of Zn$_4$Sb$_3$ have included the electronic density of states and formation energy of various atomic configurations ranging in composition from  Zn$_{12}$Sb$_{10}$ to Zn$_{14}$Sb$_{10}$\cite{Singh:Zn4Sb3:calc,Cargnoni:DOS,Toberer:Computation,Haus:MO:theory}.  These results confirm the expectation from charge counting that Zn$_{13}$Sb$_{10}$ has a Fermi level in the band gap, Zn-deficient Zn$_{12}$Sb$_{10}$ has a Fermi level in the valence band and Zn-rich Zn$_{14}$Sb$_{10}$ has a Fermi level in the conduction band.   All reports of formation energy\cite{Cargnoni:DOS,Toberer:Computation,Haus:MO:theory} also show agreement that all Zn$_4$Sb$_3$ configurations have a positive formation enthalpy at 0K with respect to Zn and ZnSb.  Yet Zn$_4$Sb$_3$ is observed below ~700K.  Furthermore, at low temperature, a reversible transition to a low symmetry, meta-stable, ordered `$\alpha$' (and 
$\alpha^{\prime}$) phase is observed\cite{Haus:alpha,Haus:alphaPrime}.  

Synthetically, Zn$_4$Sb$_3$ is usually prepared with excess Zn: Zn$_{13.3}$Sb$_{10}$.  As Zn has a high vapor pressure even at the relatively low synthesis and operation temperature, as well as the tendency to readily oxidize\cite{Pederson:Stability}, it is not very surprising that excess Zn helps form the phase.  Zn$_4$Sb$_3$ is always a heavily doped p-type semiconductor with room temperature (Hall) carrier concentration between $6-9\mathsf{x}10^{19}/$cm$^3$.\cite{Toberer:nano} According to the simple charge counting and ab initio calculations, this would be best explained by a slight Zn deficiency, Zn$_{13-\delta}$Sb$_{10}$, where $\delta$ should be between 0.016 and 0.024.  If  $n/n_{Hall}$ is ~2 as predicted by Singh\cite{Singh:Zn4Sb3:calc}, then $\delta$ should be between 0.008 and 0.012. It is perhaps surprising that with all the interstitial Zn sites, synthesis with excess Zn does not produce n-type material.

In this work, we expand on previous ab initio results to consider a thorough thermodynamic investigation of Zn$_4$Sb$_3$.  A thermodynamic ensemble approach is necessary because no single configuration (or even small number of configurations) can adequately describe the phase.  By assembling a Grand Canonical Partition Function we are able to quantitatively prove entropic stabilization.  We also predict a region of single phase stability on a temperature versus composition diagram that exhibits retrograde solubility of Zn with $\delta>0$ that would always produce p-type Zn$_4$Sb$_3$ in intermediate temperatures.

\subsection{Computational Methods}  

The natural thermodynamic function to account for a range of possible compositions is the Grand Canonical Potential\cite{avdw:emc2} (GCP):
\begin{equation}\label{gcp} 
\phi\left(T,\mu\right)=-\frac{k_BT}{N}ln\left(\sum_{s}e^{-(E_s-\mu\cdot\,N_s)/k_BT}\right)
\end{equation}
where $k$ is Boltzmann's constant, $T$ is temperature, $N$ is the total number of atoms in the system, $E_s$ is the total energy of state $s$, $N_s$ is a vector of the number of atoms in each species in state $s$ (with elements summing to $N$) and $\mu$ is a vector containing the chemical potential of each species.  This greatly simplifies for both Zn and ZnSb, which do not exhibit configurational disorder:
\begin{equation}\label{gcp2} 
\phi\left(T,\mu\right)=\frac{\varepsilon_0}{n}-\mu\cdot\,x_0
\end{equation}
where $\varepsilon_0$ is the energy per unit cell, $n$ is the number of atoms per unit cell and $x_0$ is the composition in atomic fraction.   

For Zn$_4$Sb$_3$, we consider the GCP within an "independent cells" approximation.  We assume that each primitive unit cell of $23\pm2$ atoms is non-interacting with neighboring cells, in the sense that the defect configuration present in one cell does not affect the energies of defect configurations in a nearby cell. (All our ab initio calculations are nevertheless performed on infinite periodic systems with suitable k-point sampling.)  This approach is useful in this system because the unit cell is rather large. This assumption is validated by computing the energy of supercells with different configurations in each primitive unit cell and comparing it to that predicted by summing the energies of the constituent primitive cells.  Under the independent cells approximation:
\begin{multline}\label{gcp3} 
\phi\left(T,\mu\right)=\frac{\varepsilon_0}{n}-\mu\cdot\,x_0\\
-\frac{k_BT}{n}ln\left(1+\sum_{i>0}m_ie^{-(\Delta\varepsilon_i-\mu\cdot\Delta\,n_i)/k_BT}\right)
\end{multline}
where $\varepsilon_0$ is the ground state energy per unit cell, $n$ is the number of atoms per unit cell and $x_0$ is the ground state composition.  For each configuration, $i$, $m_i$ is the symmetric multiplicity, $\Delta\varepsilon_i$ is the change in energy from the ground state and $\Delta\,n_i$ is the change in the number of atoms from the ground state. 

The energy for each configuration is calculated under the generalized gradient approximation (GGA) using the projector augmented wave (PAW) method with Perdew-Burke-Ernzerhof (PBE) potentials as implemented in VASP 4.6, neglecting spin orbit coupling.  All unit cell parameters and atomic positions were allowed to relax to find the lowest energy configuration to within $10^{-4}$ eV.  A final static calculation was performed for an accurate total energy.

Defect configurations were systematically generated by enumerating defect combinations deviating from the undefected Zn$_{12}$Sb$_{10}$ structure (`A' sub-lattice fully occupied and `B',`C and `D' sub-lattices fully unoccupied).  Defect clusters were composed of a combination of vacancies on the `A' sub-lattice, and occupation of an interstitial site on the `B', `C' or `D' sub-lattices.  Clusters of up to 6 defects were enumerated, excluding some structures because the defects were too close (nearest neighbor A and C sites both occupied).  After allowing the atomic configuration to relax, the resulting atomic positions were projected onto the closest unrelaxed configuration and any duplicate configurations were excluded in order to avoid over-counting states. 

Phonon density of states and vibrational free energies were calculated using the `supercell' method as implemented in the Alloy Theoretic Automated Toolkit (ATAT)\cite{avdw:atat,avdw:atat2,avdw:maps}.  Since the computational resources needed to compute phonon modes for all Zn$_4$Sb$_3$ configurations are prohibitive, representative configurations were selected at four compositions between Zn$_{12}$Sb$_{10}$ and Zn$_{15}$Sb$_{10}$ configurations.  A second Zn$_{13}$Sb$_{10}$ configuration was computed as well to assess the error in our approximation.  For each of these configurations and the end-members Zn and ZnSb, phonon modes were calculated at 0\%, 1\% and 2\% strain to account for the effects of thermal expansion, under the quasi-harmonic approximation.

The vibrational contribution to the free energy is incorporated into the GCP through a nested sum in the partition function\cite{avdw:vibrev}.  For each distinct configuration, phonon occupation accounts for small displacements around the local energy minimum, resulting in a temperature dependent free energy correction. 

GCPs were assembled for Zn, Zn$_4$Sb$_3$ and ZnSb.  Phase equilibrium is determined by equality of two respective GCPs.  The equilibrium composition of each phase can be determined by:
\begin{equation}\label{gcp4} 
\nabla_\mu\phi\left(T,\mu\right)=-x
\end{equation}

\subsection{Results and Discussion}

\begin{figure}
\epsfig{file=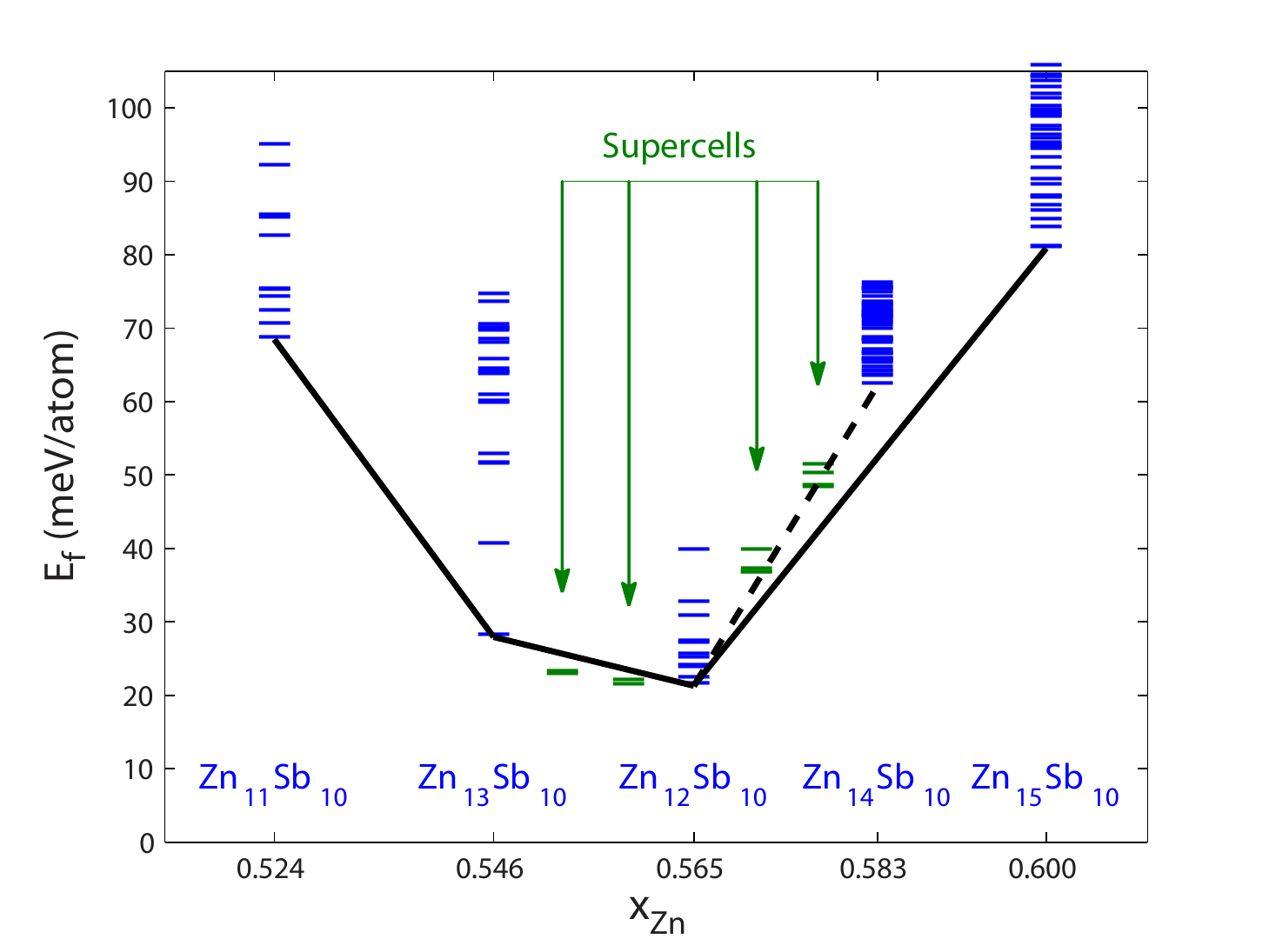, width=6.5cm}\\
\caption{(Color online) Formation enthalpy at 0K for Zn$_4$Sb$_3$ configurations with respect to Zn and ZnSb.  The solid black line represents the convex hull of structures confined to the Zn$_4$Sb$_3$ lattice.  Supercell calculations in green were done using the conventional unit cell, each composed of three primitive unit cells. The dashed line shows the predicted energy for supercells following the independent cells approximation. }
\label{FormEnergy}
\end{figure}

Over 100 unique, stable configurations were enumerated in the $\beta-$Zn$_4$Sb$_3$ primitive rhombohedral unit cell with between 21 and 25 atoms per cell (the number of Sb was held constant at 10).  The 0K formation enthalpy of these configurations with respect to Zn and ZnSb is shown in Figure\,\ref{FormEnergy}.

Consistent with previous theoretical studies\cite{Cargnoni:DOS,Toberer:Computation,Haus:MO:theory}, all configurations have a positive formation enthalpy with respect to Zn and ZnSb.  The "convex hull" is plotted in solid black and connects the ground state configurations (for Zn$_4$Sb$_3$ phase) at each composition.  The composition Zn$_{14}$Sb$_{10}$ does not have a configuration touching the convex hull.  This means that for an ensemble of atoms of composition Zn$_{14}$Sb$_{10}$ constrained to remain in the Zn$_4$Sb$_3$ lattice, it would be more energetically favorable to form a mixture of cells of composition Zn$_{13}$Sb$_{10}$ and Zn$_{15}$Sb$_{10}$.  If the system is allowed to adopt any lattice, then, at 0K, it would be even more energetically favorable to separate into Zn and ZnSb crystals.  Figure\,\ref{FormEnergy} also allows us to test the independent cells approximation.  The data points marked "supercells" are all made up of 3 primitive cells of different compositions to yield the intermediate compositions shown.  Between Zn$_{12}$Sb$_{10}$ and Zn$_{13}$Sb$_{10}$, we see that the supercells produce a slightly lower energy by about 3\,meV/atom than predicted by the independent cells approximation.  Between Zn$_{13}$Sb$_{10}$ and Zn$_{14}$Sb$_{10}$, the energies from the supercells are either very close or slightly higher than predicted by the independent cells approximation (the dashed line).  The small energy difference of a few meV/atom justifies the use of the independent cells approximation, although a consideration of a possible systematic error is discussed later.  

\begin{figure}
\epsfig{file=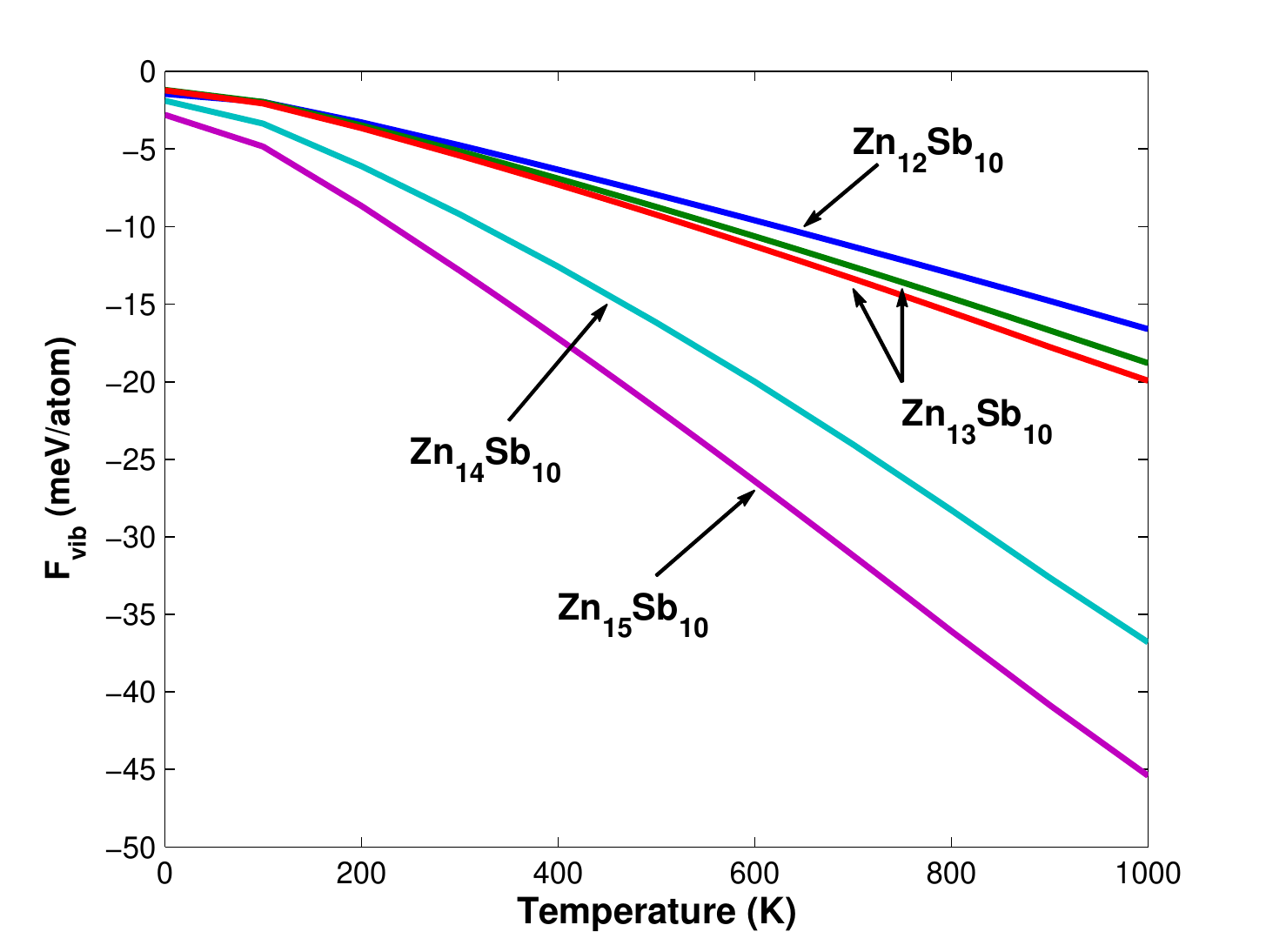, width=6.5cm}\\
\caption{Phonon contribution to the formation free energy with respect to Zn and ZnSb.  The two Zn$_{13}$Sb$_{10}$ configurations are in good agreement to ~1\,meV/atom.   There is a more favorable contribution to the formation energy for more disordered Zn-rich configurations.  }
\label{Fvib}
\end{figure}

The formation free energy due to phonons (relative to Zn and ZnSb) is shown in Figure\,\ref{Fvib} for the 5 representative configurations.  In all cases, the phonon contribution to the free energy favors the Zn$_4$Sb$_3$ phase over Zn and ZnSb.  This is likely a result of the softer phonon modes in the more complex and open crystal structure.  Increasing Zn concentration and thereby increasing the Zn-disorder results in a more favorable contribution to the free energy.  There is good agreement between the two configurations of composition Zn$_{13}$Sb$_{10}$, differing by ~1\,meV/atom at 1000K.  It then seems reasonable to assume that the vibrational free energy of the representative structures may be applied to all the configurations at that composition when we compute the GCP of Zn$_4$Sb$_3$.      

\begin{figure}
\epsfig{file=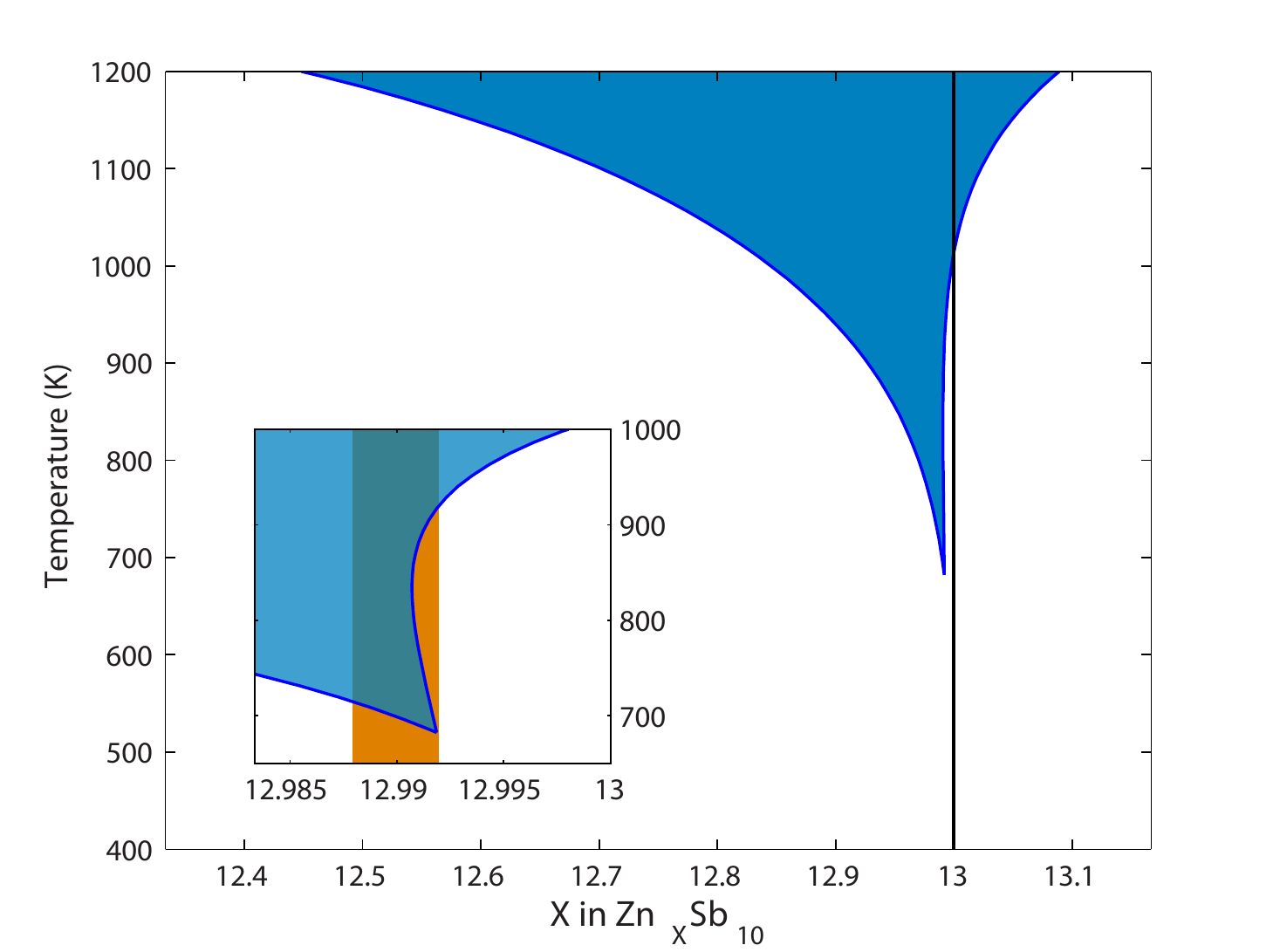, width=6.5cm}\\
\caption{(Color online) Single phase stability region for Zn$_4$Sb$_3$.  Solid black line corresponds to the charge balanced composition.  Stable compositions to the left of the line will result in a p-type semiconductor.  Inset shows a  close-up of the region of retrograde Zn solubility.  The orange band corresponds to compositions estimated from charge carrier measurements\cite{Toberer:nano}. }
\label{PhaseDiagram}
\end{figure}

The computed region of single phase stability for Zn$_4$Sb$_3$ is shown in Figure\,\ref{PhaseDiagram}.  Including configurational and vibrational effects to the free energy, Zn$_4$Sb$_3$ is found to stabilize at around 700K at a composition of $Zn_{12.992}Sb_{10}$.  As the temperature increases, the range of stable single phase compositions increases, more broadly on the Zn deficient side.  (Eventually, the solid melts, but our analysis focuses on the solid-state portion of the phase diagram.)  Interestingly, on the Zn-rich side, we observe retrograde solubility of Zn.   That is, the highest stable concentration of Zn in Zn$_4$Sb$_3$ decreases with increasing temperature.  This unusual finding is a result of the composition-dependent relationship between the enthalpy and entropy of mixing. Typically defects of all kinds entropically stabilize stoichiometric variations on both sides of a valence precise composition. In the case of Zn$_4$Sb$_3$, the valence precise composition is already `defected' with interstitial Zn.  A miscibility gap arises for compositions above Zn$_{13}$Sb$_{10}$ due to the large jump in formation enthalpy compared to the Zn-deficient compositions.  The effect is lessened by the counteractive effects of higher vibrational entropy for the Zn-rich compositions.  If the vibrational effects were neglected, the Zn-rich boundary would not curve to the right until a much higher temperature.  
 
The region of single phase stability shown in Figure\,\ref{PhaseDiagram} has readily observable consequences.  The black vertical line at composition Zn$_{13}$Sb$_{10}$ represents the valence precise structure with a Fermi level inside the band gap.  At any Zn-deficient composition from Zn$_{13}$Sb$_{10}$, we expect a partially filled valence band resulting in a p-type semiconductor.  (Surprisingly, there is a recent report of a Zn$_{13}$Sb$_{10}$ configuration being slightly p-type\cite{Wrong:DOS}.  Our electronic density of states results are in agreement with previous studies\cite{Singh:Zn4Sb3:calc,Cargnoni:DOS,Toberer:Computation,Haus:MO:theory}.)  All stable compositions below 1000K in Figure\,\ref{PhaseDiagram} result in a p-type semi-conductor.   Attempting to dope Zn$_4$Sb$_3$ with excess Zn will not result in an n-type semiconductor but a two-phase equilibrium between Zn metal and p-type Zn$_4$Sb$_3$.  Furthermore, with the consideration of retrograde solubility, if Zn-saturated single phase Zn$_4$Sb$_3$ is heated up, it passes through a region where it becomes thermodynamically more stable to precipitate Zn metal.  With more heating, the Zn is reabsorbed as more Zn-rich compositions are stable.  Upon cooling, the same precipitation and absorption should occur if held in thermodynamic equilibrium.  

On the Zn-deficient side of the stable phase region the usual temperature dependent solubility is observed. Here one would expect ZnSb to precipitate as the temperature is reduced. Because this would occur at low temperature in the solid state, small nanometer sized precipitates would be expected. Such particles are indeed observed in Sb-rich samples \cite{Toberer:nano}.

The high temperature phase boundary on the Zn-rich side could explain the formation of Zn nanoparticles observed in some Zn$_4$Sb$_3$ samples \cite{Pryzt:nano}.  Cooling a Zn-rich composition above Zn$_{13}$Sb$_{10}$ would precipitate Zn (possibly nano-particles if cooled fast enough\cite{Ikeda:Precipitates}). Upon further cooling through the retrograde region, some of the Zn would be reabsorbed into the $\beta-$phase.  The absorption of nanoparticle Zn may explain the nano-voids observed in some samples \cite{Toberer:nano}

Experimental carrier concentration measurements from Toberer\cite{Toberer:nano} from single phase $\beta-$Zn$_4$Sb$_3$ samples range from $6-9\mathsf{x}10^{19}/$cm$^3$.  This corresponds to a composition range of 0.002\% atomic Zn between Zn$_{12.988}$Sb$_{10}$ and Zn$_{12.992}$Sb$_{10}$ assuming $n/n_{Hall}=2$.\cite{Singh:Zn4Sb3:calc}.  This concentration range, corresponding to the orange band in the inset of Figure\,\ref{PhaseDiagram}, is in close agreement with the predicted stable compositions near the stabilization temperature.   This composition range is much smaller than the 0.2\% atomic Zn observed by microprobe analysis.  This discrepancy is not surprising since the composition range is basically the limit of the microprobe resolution. 

There is some expected uncertainty in our predicted temperature of stabilization of Zn$_4$Sb$_3$.  Our predicted temperature of 700K is much higher than might be expected. There are several possible explanations for this.  One could be a deficiency in our methodology to fully account for the sources of entropy in such a complex disordered structure.  Our enumeration method may underestimate the number of Zn$_4$Sb$_3$ configurations by comparing the relaxed structures to the fixed lattice in Figure\,\ref{structure}, which represents, in a sense, a measured average over many possible configurations.  The vibrational contribution to the free energy has a significant effect in lowering the stabilization temperature.  Assuming a representative structure for each composition might be an underestimate for the more disordered configurations and the effect could be even more pronounced.  Lastly, we noted earlier that the supercells on the Zn-deficient side of Zn$_{13}$Sb$_{10}$ had slightly lower formation energy than expected under the independent cells approximation.  This neglected cell-to-cell interaction (~2\,meV/atom) could lower the stabilization temperature by several hundred degrees. In either case, we expect the retrograde solubility to remain, and possibly intensify, because the supercells on the Zn-rich side of Zn$_{13}$Sb$_{10}$ are in good agreement with the independent cells approximation.  

To ensure that our finding of a single-phase region for the $\beta$-Zn$_4$Sb$_3$ lattice is not an artifact of the independent cell approximation, we have conducted separate Monte Carlo simulations based on a cluster expansion Hamiltonian fitted to our database of structural energies. These simulations confirm the presence of a single phase region over the temperature range where $\beta$-Zn$_4$Sb$_3$ is stable. Hence, the independent cell approximation was deemed reliable and was used throughout this work. It provides a convenient explicit expression for the free energy and is immune to the fitting errors inherent to the Hamiltonian construction procedure.

Finally, we consider how Zn$_4$Sb$_3$ might interact with other nearby phases, namely, $\alpha-$Zn$_4$Sb$_3$ and Zn$_8$Sb$_7$.  First principles calculation of $\alpha-$Zn$_4$Sb$_3$ yields positive formation energy of 19\,meV/atom with respect to Zn and ZnSb.  This is 3\,meV/atom below the lowest energy $\beta-$Zn$_4$Sb$_3$ configuration of composition Zn$_{13}$Sb$_{10}$.   Assuming no configurational disorder in $\alpha-$Zn$_4$Sb$_3$ we predict $\beta-$Zn$_4$Sb$_3$ to become energetically favorable comparted to $\alpha-$Zn$_4$Sb$_3$ at ~300K (although both are still meta-stable with respect to Zn and ZnSb at this temperature).   It seems possible that there is a temperature range in which meta-stable $\beta-$Zn$_4$Sb$_3$ is observed before  becoming thermodynamically stable with respect to Zn and ZnSb at some higher temperature. 

\begin{figure}
\epsfig{file=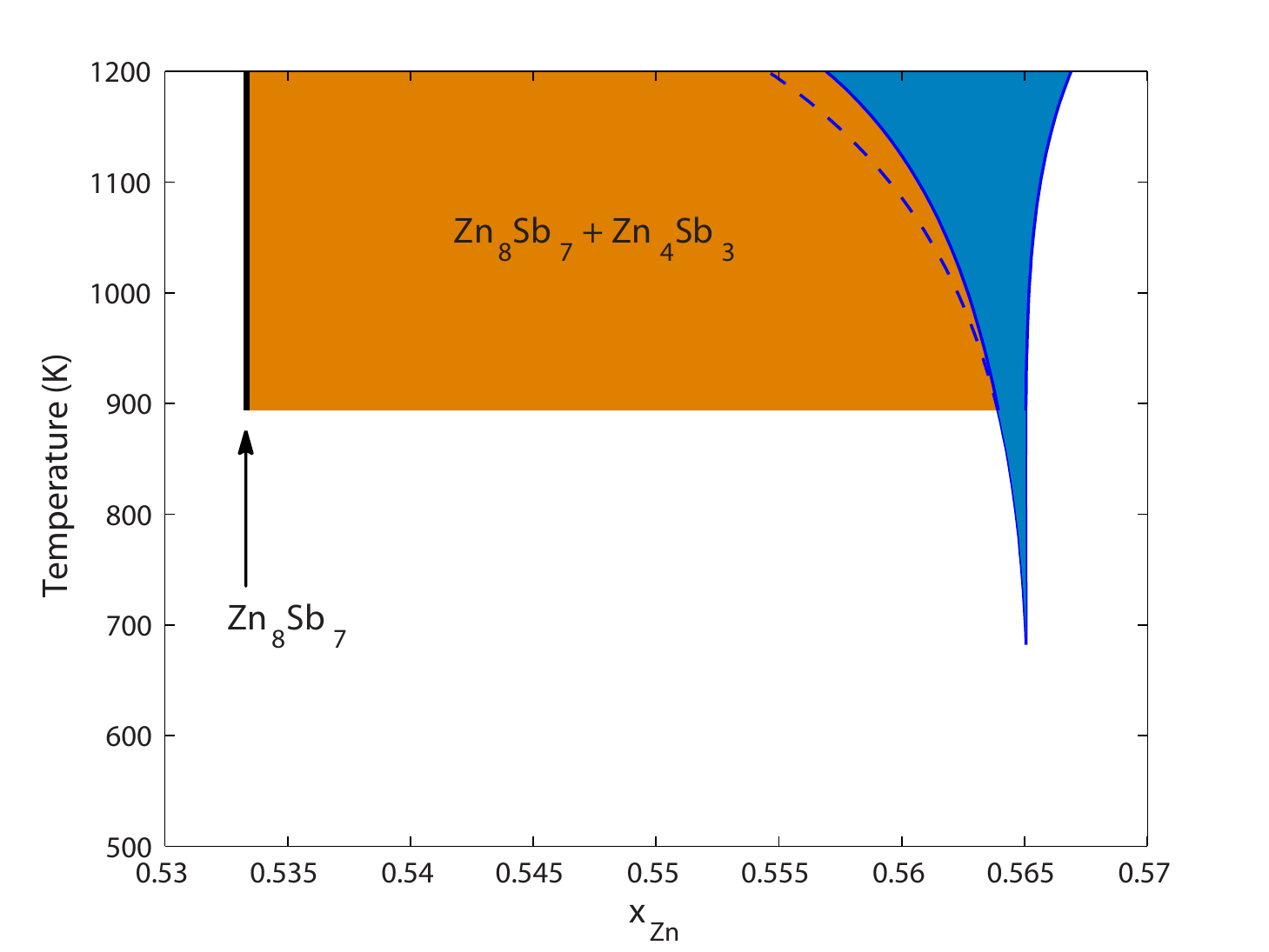, width=6.5cm}\\
\caption{(Color online) Possible phase stability regions of Zn$_4$Sb$_3$ and Zn$_8$Sb$_7$ under conditions favorable to the formation of Zn$_8$Sb$_7$.  The dotted line represents the phase boundary of Zn$_4$Sb$_3$ in the absence of Zn$_8$Sb$_7$.  }
\label{PhaseDiagram2}
\end{figure}

Another phase of recent interest is Zn$_8$Sb$_7$, which has been characterized experimentally and studied through first principles calculations \cite{Chris:Zn8Sb7,Pomrehn:Zn8Sb7}.   These calculations reveal that, even though Zn$_8$Sb$_7$ is unstable in bulk form, it could be stabilized in nanocrystaline form if surface stress or energy contributions result in a free energy decrease of ~5\,meV/atom (relative to ZnSb and Zn$_4$Sb$_3$).  Taking these assumptions into effect, the resulting phase diagram is shown in Figure\,\ref{PhaseDiagram2}.  We see that Zn$_8$Sb$_7$ would stabilize at a temperature higher than Zn$_4$Sb$_3$ and a two phase region results. This also reduces the stable Zn-deficient Zn$_4$Sb$_3$ composition range predicted in the absence of Zn$_8$Sb$_7$ (dashed line in Figure\,\ref{PhaseDiagram2}). Other unidentified phases as reported in some phase diagrams\cite{Izard:phase:diagram} could have a similar effect.   

\subsection{Conclusions}
From our first principles investigation, we have shown that Zn$_4$Sb$_3$ is entropically stabilized with the help of configurational and vibrational entropy.  Under the independent cells approximation of the grand canonical potential we predict a region of single phase stability near Zn$_{12.992}$Sb$_{10}$, which results in a nominally p-type semi-conductor.  Additionally, we predict a temperature range with retrograde Zn solubility.  The temperature dependent solubility can be used to explain the variety of nanoparticle formation observed in the system: formation of ZnSb on the Sb-rich side, Zn on the far Zn-rich side and nano-void formation due to Zn precipitates being reabsorbed at lower temperatures. 

\subsection*{Acknowledgements}
This work is supported by the U.S. National Science Foundation via grant DMR-0953378 and via TeraGrid resources at NCSA and SDSC under Grant No. TG-DMR050013N.  The authors also thank the Caltech MURI project and DARPA. 

\bibliography{Pomrehn_Zn4Sb3_final1}

\end{document}